\def\centeron#1#2{{\setbox0=\hbox{#1}\setbox1=\hbox{#2}\ifdim
\wd1>\wd0\kern.5\wd1\kern-.5\wd0\fi
\copy0\kern-.5\wd0\kern-.5\wd1\copy1\ifdim\wd0>\wd1
\kern.5\wd0\kern-.5\wd1\fi}}
\def\centerover#1#2{\centeron{#1}{\setbox0=\hbox{#1}\setbox
1=\hbox{#2}\raise\ht0\hbox{\raise\dp1\hbox{\copy1}}}}
\def\centerunder#1#2{\centeron{#1}{\setbox0=\hbox{#1}\setbox
1=\hbox{#2}\lower\dp0\hbox{\lower\ht1\hbox{\copy1}}}}
\def\lsim{\;\centeron{\raise.35ex\hbox{$<$}}{\lower.65ex\hbox
{$\sim$}}\;}
\def\gsim{\;\centeron{\raise.35ex\hbox{$>$}}{\lower.65ex\hbox
{$\sim$}}\;}

\documentclass{ws-ijmpa}

\begin{document}

\markboth{Geoffrey T.~Bodwin}{NRQCD: Fundamentals and Applications}

\catchline{}{}{}{}{}

\title{NRQCD: Fundamentals and Applications to Quarkonium Decay and 
Production}

\author{Geoffrey T.~Bodwin}
\address{HEP Division, Argonne National Laboratory, 9700 South Cass
Avenue, Argonne, IL 60439}


\maketitle


\begin{abstract}
I discuss NRQCD and, in particular, the NRQCD factorization formalism for
quarkonium production and decay. I also summarize the current status of
the comparison between the predictions of NRQCD factorization and
experimental measurements.
\end{abstract}



\section{Nonrelativistic QCD (NRQCD)}

In a heavy quarkonium, a bound state of a heavy quark and antiquark,
there are many important momentum scales. These include $m$, the
heavy-quark mass; $mv$, the typical heavy-quark momentum in the
quarkonium rest frame; and $mv^2$, the typical heavy-quark rest-frame
kinetic energy and binding energy. Here $v$ is the typical heavy-quark
velocity in the quarkonium rest frame. For charmonium, $v^2\approx 0.3$,
while for bottomonium, $v^2\approx 0.1$.

In theoretical analyses, it is useful to treat the physics at each of
these scales separately. Owing to asymptotic freedom, interactions at
the scale $m$ can be treated perturbatively. Approximate symmetries
({\it e.g.} heavy-quark spin symmetry) can be exploited at some scales.
Often, analytic calculations simplify when they involve only one
scale at a time. Lattice calculations can encompass only a
limited range of scales, and so become more tractable after scale
separation.

Effective field theories provide a convenient way to separate scales. An
effective theory describes the low-momentum degrees of freedom in the
original theory. It is constructed by integrating out the high-momentum
degrees of freedom in that theory. Nonrelativistic QCD (NRQCD) is an
effective field theory that separates scales of order $m$ and higher
from the other scales in
QCD.\cite{Caswell:1985ui,Thacker:1990bm,Lepage:1992tx} NRQCD has a UV
cutoff $\Lambda\sim m$. For processes with $p<\Lambda$, NRQCD reproduces
QCD. Processes with $p>\Lambda$ are not manifest in NRQCD, but they
affect the coefficients of local interactions. $\Lambda$ plays the
r\^ole of a factorization scale between the hard and soft physics.

At leading order in $v$, the NRQCD action is just the Schr\"odinger-Pauli
action for a heavy quark and a heavy antiquark. In order to reproduce QCD
completely, one would need an infinite number of interactions of all
orders in $v$. However, in practice one works to a given precision in $v$.

NRQCD and NRQED (the corresponding effective theory for QED) have had a
number of well-established successes. These include calculations of the
properties of QED bound states, such as positronium and muonium, lattice
calculations of quarkonium spectra, and lattice calculations of
$\alpha_s$. However, here we will focus on recent applications to
quarkonium decay and production.

\section{Quarkonium Decay and Production}

In heavy-quarkonium decays and hard-scattering production, large scales
appear: Both the heavy-quark mass $m$ and the quarkonium transverse
momentum $p_T$ are much larger than $\Lambda_{\rm QCD}$. The hope is
that NRQCD would allow one to separate these short-distance,
perturbative scales from the long-distance quarkonium dynamics.

In the case of quarkonium decays, convincing arguments have been given 
that the short-distance decay process can be represented in NRQCD by
point-like four-fermion interactions.\cite{BBL} This result leads to the 
factorization formula
\begin{equation}
\Gamma(H \to \hbox{light hadrons})
\;=\; \sum_n 2 \; {\rm Im \,} F_n(\Lambda)\;
        {\langle H |} {\cal O}_n(\Lambda) {| H \rangle} .
\label{decay-fact}
\end{equation}
The $F_n$ are ``short-distance coefficients,'' which are determined by
perturbative matching of amplitudes between QCD and
NRQCD. The factors ${\langle H |} {\cal O}_n(\Lambda) {| H \rangle}$ are
inherently nonperturbative matrix elements of four-fermion operators in
the quarkonium state. The sum in Eq.~(\ref{decay-fact}) is actually an
expansion in powers of $v$ and is truncated at some finite order. A
similar factorization formula has been conjectured to hold for inclusive
quarkonium production at large $p_T$.\cite{BBL} The production matrix
elements are the crossed versions of the decay matrix elements. Only the
color-singlet production and decay matrix elements are simply related.
Nayak, Qiu, and Sterman have found that, in next-to-next-to-leading
order in $\alpha_s$, the NRQCD production matrix elements must be
modified by the inclusion of eikonal lines in order to allow one to
factor all of the IR-divergent contributions out of the short-distance
coefficients.\cite{Nayak:2005rw} This finding does not affect existing
phenomenology, which is at the tree level or one-loop level. However, it
raises important issues with regard to the validity of the NRQCD
factorization formula at all orders in $\alpha_s$.

The NRQCD factorization formalism gains much of its predictive power
from the fact that the nonperturbative matrix elements are universal,
{\it i.e.}, process independent. Although some decay matrix elements
have been computed on the lattice,\cite{lattice} in
general, the matrix elements must be extracted phenomenologically. The
consistency of the phenomenological matrix elements from process to
process is a key test of the NRQCD factorization formalism.

An important feature of the NRQCD factorization formalism is that
quarkonium decay and production occur through color-octet, as well as
color-singlet, $Q\overline Q$ states.  If one drops all of the
color-octet contributions, then the result is the color-singlet model
(CSM).

\section{Comparison of NRQCD factorization with experiment}

\subsection{$P$-wave decays}

A global fit of the NRQCD predictions for the inclusive decay rates of
$P$-wave quarkonia (Ref.~\refcite{Maltoni:2000km}) is in good overall
agreement with the data and yields values for the NRQCD matrix elements
that are in good agreement with those from lattice calculations
(Ref.~\refcite{lattice}). More recently, a number of
next-to-leading-order NRQCD predictions for decay rates of $P$-wave
quarkonia have been verified by more precise experimental
measurements.\cite{Vairo:2004sr}

\subsection{Quarkonium production at the Tevatron}

\begin{figure}
\centerline{\includegraphics[width=6.5cm]{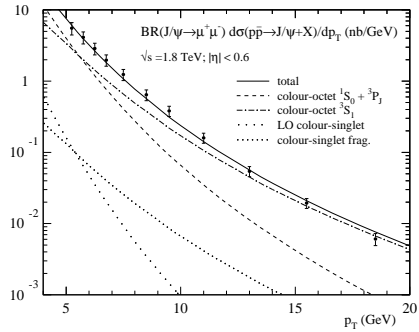}}
\caption{The $J/\psi$ cross section as a function of $p_T$. The data
points are from the CDF measurement.\protect\cite{Abe:1997jz} The solid curve
is the NRQCD factorization fit to the data given in
Ref.~\protect\refcite{Kramer:2001hh}. The other curves give various 
contributions
to the NRQCD factorization fit. From
Ref.~\protect\refcite{Kramer:2001hh}.}
\label{fig:tevatron-j-psi}
\end{figure}

The CDF data for $J/\psi$ production at the Tevatron are shown in
Fig.~\ref{fig:tevatron-j-psi}, along with the NRQCD factorization result.
As can be seen, the color-singlet contributions, whose normalizations
are reasonably well known from decay processes, are smaller than the data
by more than an order of magnitude. The color-octet contributions
bring the theory into good agreement with the data. However, it should
be remembered that the color-octet matrix elements are obtained by
fitting to the Tevatron data. The shape of the data is consistent with
NRQCD factorization, but there is a good deal of freedom to change the
predicted shape by adjusting the relative values of the color-octet matrix
elements. The Tevatron data for $\psi'$ and $\Upsilon$ production are
also fit well by the NRQCD factorization expressions. In subsections
that follow, we discuss more stringent tests of NRQCD factorization, in
which the values of the color-octet matrix elements that were obtained
in the fits to the Tevatron data are used to make predictions for other
processes.

\subsection{$\bm{\gamma \gamma\rightarrow J/\psi +X}$ at LEP}

\begin{figure}
\centerline{\includegraphics[width=6.0cm]{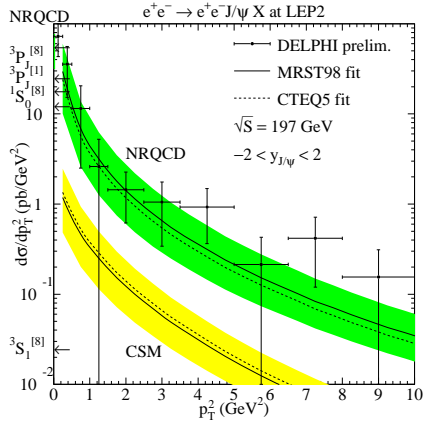}}
\caption{The cross section for ${\gamma \gamma\rightarrow J/\psi +X}$ at
LEP. The upper set of curves is
the NRQCD factorization prediction, and the lower set of curves is the
color-singlet-model prediction. The data points are from the Delphi 
Collaboration measurement.\protect\cite{delphi} From
Ref.~\protect\refcite{klasen-kniehl-mihaila-steinhauser}.\label{fig:delphi}}
\end{figure}

As can be seen in Fig.~\ref{fig:delphi}, in the case of $J/\psi$
production in $\gamma\gamma$ collisions at LEP, comparison of theory
with Delphi data clearly favors NRQCD factorization over the color-singlet
model.\cite{klasen-kniehl-mihaila-steinhauser}
The theoretical uncertainties that are shown arise from uncertainties in
the color-octet matrix elements and in the choices of renormalization
and factorization scales. The uncertainties in the color-octet matrix
elements are quite large because different linear combinations of matrix
elements appear in $J/\psi$ production at LEP than in $J/\psi$
production at the Tevatron.

\subsection{Quarkonium production in DIS at HERA}

The leading-order NRQCD factorization and CSM predictions 
(Ref.~\refcite{Kniehl:2001tk})
for the $J/\psi$ inclusive production cross
sections $d\sigma/dp_T^2$ and $d\sigma/dQ^2$ in $ep$ deep-inelastic 
scattering (DIS) are
in good agreement with the H1 data (Ref.~\refcite{h1-dis-psi}).
Those data lie well above the CSM prediction.
The cross section $d\sigma/dz$, which is differential in
the energy fraction (inelasticity) $z$, is not fit well by either the
NRQCD factorization or CSM predictions. The poor fit of the NRQCD
factorization prediction may be a consequence of the breakdown of
both the $v$ expansion and the $\alpha_s$ expansion near $z=1$. This
phenomenon will be discussed in conjunction with inelastic quarkonium
photoproduction at HERA below. The ZEUS data for $d\sigma/dQ^2$ agree
less well with the NRQCD prediction than the H1 data, but they have
larger error bars.\cite{Katkov:2003gj}

\subsection{Polarization of quarkonia at the Tevatron}

\begin{figure}
\centerline{\includegraphics[width=5.0cm]{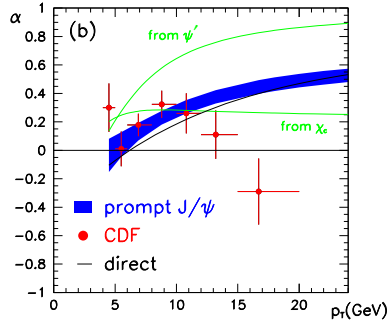}}
\caption{$J/\psi$ polarization at the Tevatron. $\alpha=1$ corresponds
to 100\% transverse polarization;  $\alpha=-1$ corresponds to 100\%
longitudinal polarization. The band is the total NRQCD factorization
prediction. The other curves give the contributions from feeddown from
higher charmonium states. The data points are from the CDF
measurement.\protect\cite{Affolder:2000nn} From
Ref.~\protect\refcite{braaten-kniehl-lee-pol}.\label{fig:psi-pol}}
\end{figure}

The polarizations of quarkonia produced at large $p_T$ at the Tevatron
provide potentially definitive tests of the color-octet mechanism.
$J/\psi$ production at large-$p_T$ is expected to be dominated by gluon
fragmentation into a color-octet $Q\overline Q$ pair. This mechanism
leads to transversely polarized $J/\psi$'s.\cite{cho-wise,YR} The NRQCD
factorization prediction for the $J/\psi$ polarization as a function of
$p_T$ (Ref.~\refcite{braaten-kniehl-lee-pol}) is shown, along with the
CDF data, in Fig.~\ref{fig:psi-pol}.
The observed $J/\psi$ polarization is generally smaller than the
prediction and seems to trend in the wrong direction, decreasing with
increasing $p_T$.  However, the experimental error bars are large, and
only the last data point truly disagrees with the prediction. There are
large uncertainties in the theoretical predictions that arise from
uncertainties in the NRQCD matrix elements, which are indicated in the
prediction band in Fig.~\ref{fig:psi-pol}. There are also large
corrections of higher order in $\alpha_s$ and $v$ to the quarkonium
production rate, some of which have been calculated.\cite{YR} To first
approximation, such corrections merely change the normalizations of the
fitted color-octet matrix elements without changing the polarization
strongly. It has been suggested that nonperturbative spin-flip
processes, which are suppressed as $v^3$ and are not taken into account
in present calculations, might be important.\cite{YR} However, a recent
lattice calculation suggests that this is not the
case.\cite{Bodwin:2005gg}

\subsection{Inelastic photoproduction at HERA}

\begin{figure}[htb]                                             
\begin{tabular}{cc}
\includegraphics[width=6.5cm]{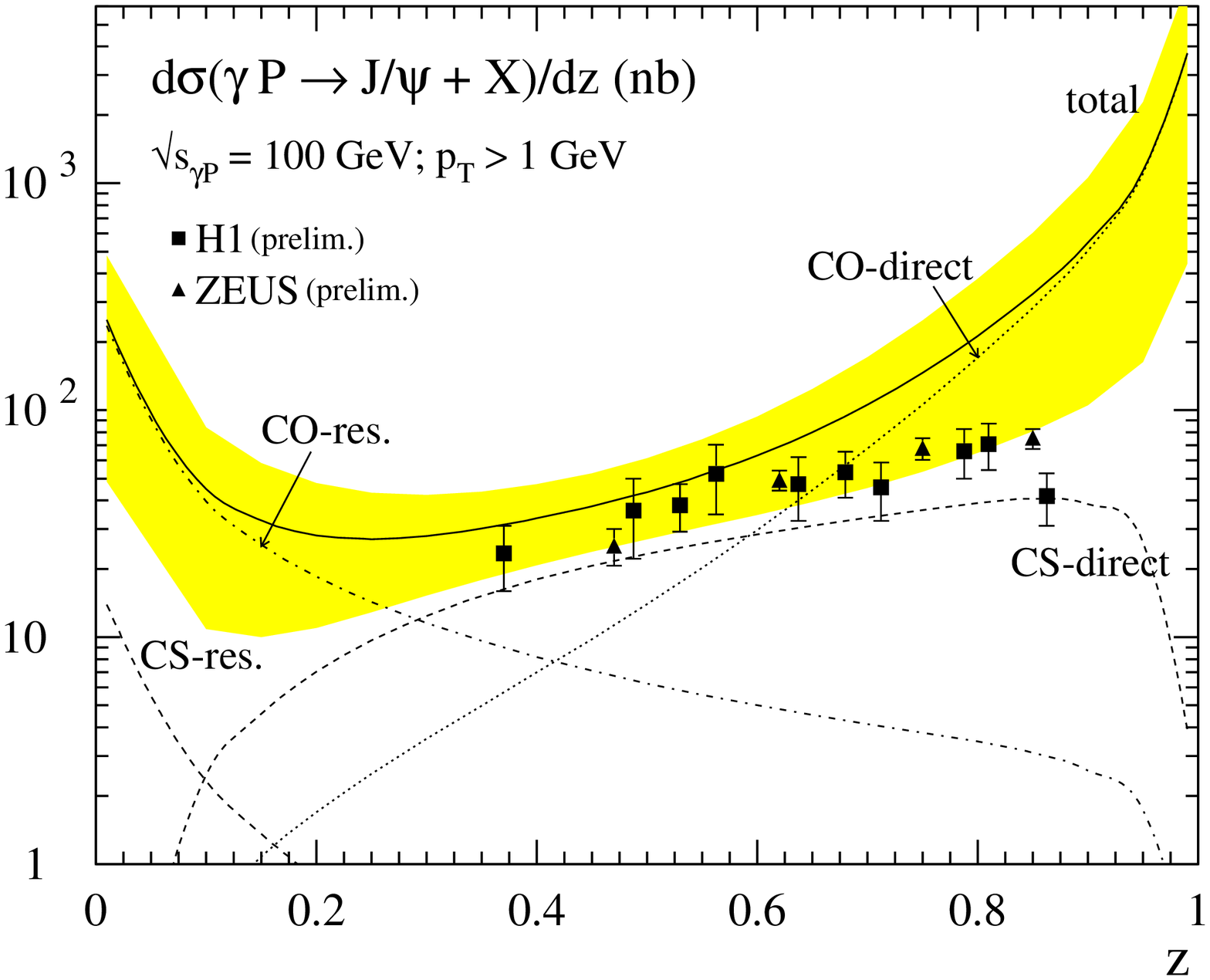}
&\hbox{}\hskip -0.5cm\includegraphics[width=6.5cm]{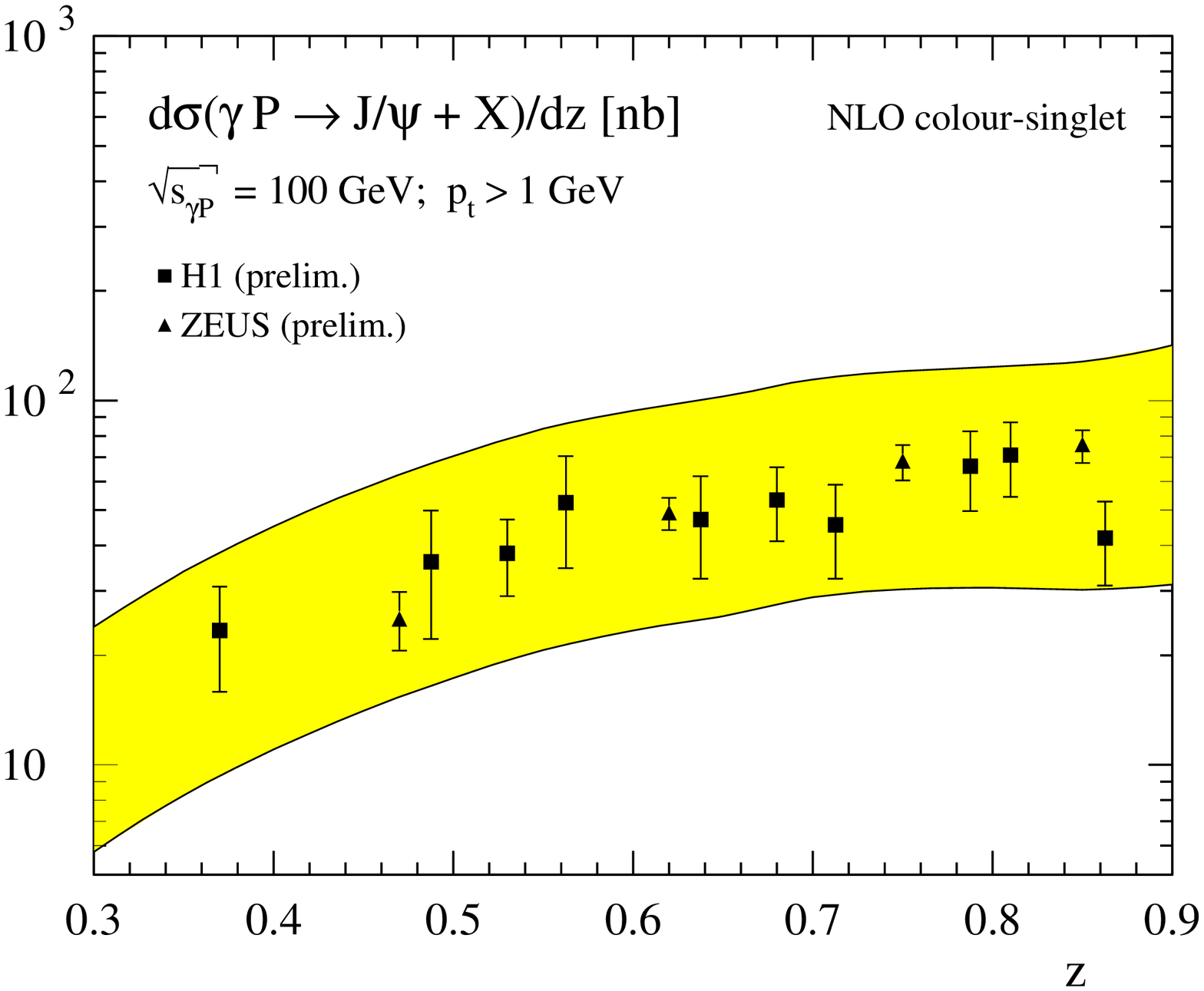}
\end{tabular}
\caption{The rate for inelastic quarkonium photo-production at HERA as a
function of the energy fraction $z$. In the left-hand figure, the curves
give leading-order (LO) total, direct, and resolved color-octet (CO) and
color-singlet (CS) contributions. The band shows the uncertainty in the
total contribution that arises from the uncertainties in the color-octet
matrix elements. In the right-hand figure, the curves show the
prediction in NLO of the color-singlet model. The band shows the
uncertainties that arise from $\alpha_s$ and $m$. The data points are
from the H1 and Zeus results of
Refs.~\protect\refcite{H1-photoproduction} and
\protect\refcite{Zeus-photoproduction}. From
Ref.~\protect\refcite{Kramer:2001hh}.\label{fig:photoproduction}}
\end{figure}

Theoretical calculations of the cross section for inelastic
photoproduction of quarkonium at HERA have been carried out in the NRQCD
factorization formalism by several groups.\cite{YR} The compilation of
predictions from Ref.~\refcite{Kramer:2001hh} and the H1 and Zeus data
are shown in Fig.~\ref{fig:photoproduction}, plotted as function of the
energy fraction $z$.
As can be seen, the color-octet contribution is poorly determined, owing
to large uncertainties in the color-octet matrix elements. Even so,
there is little room for a color-octet contribution. Furthermore, as is
shown in Fig.~\ref{fig:photoproduction}, corrections of next-to-leading
order in $\alpha_s$ (NLO) increase the color-singlet piece by about a
factor of two at large $z$. The color-singlet piece is then, by itself,
in good agreement with the data.\cite{Kramer:1994zi,Kramer:1995nb} The
data differential in $p_T$ are also compatible with NLO color-singlet
production alone at large $p_T$.\cite{Kramer:2001hh} It should be noted,
though, that there are large uncertainties in the NLO color-singlet
contribution, which arise primarily from uncertainties in $m_c$ and
$\alpha_s$. The true color-singlet contribution could be lower than the
central value by about a factor of two, leaving more room for a
color-octet contribution.

Near $z=1$, the leading-order color-octet contribution grows rapidly, in
apparent disagreement with the data. However, in this region, soft-gluon
emission leads to large logarithms of $1-z$ and also to large
corrections of higher order in $v$, both of which must be resummed. The
resummation of the corrections of higher order in $v$ leads to a
nonperturbative ``shape function.''\cite{Beneke:1997qw} Both the shape
function and the resummed logarithmic corrections significantly smear
out the color-octet contribution near $z=1$ and may lead to a
considerable improvement in the agreement of the NRQCD factorization
predictions with the data. 

\subsection{Double $\bm{c\overline{c}}$ production at Belle}

For the exclusive double charmonium process $e^+e^-\to J/\psi+\eta_c$,
the cross section times the branching ratio into at least two charged tracks
has been measured by the Belle collaboration to be $25.6\pm 2.8\pm
3.4~\hbox{fb}$ (Ref.~\refcite{Pakhlov:2004au}) and by the BaBar
collaboration to be $17.6\pm 2.8 {}^{+1.5}_{-2.1}~\hbox{fb}$
(Ref.~\refcite{Aubert:2005tj}). In contrast, leading-order NRQCD 
factorization calculations predict a cross section of $3.78\pm
1.26~\hbox{fb}$.\cite{Braaten:2002fi,Liu:2002wq} A similar
disagreement between NRQCD factorization and experiment holds for
production of $\chi_{c0}$ and $\eta_c(2S)$ mesons in conjunction with a
$J/\psi$ meson. A recent calculation of corrections of next-to-leading
order in $\alpha_s$ leads to an enhancement of the theoretical
prediction of about a factor 1.8 (Ref.~\refcite{Zhang:2005ch}).
Calculations in the light-cone formalism are in reasonably good
agreement with the data.\cite{Ma:2004qf,Bondar:2004sv} The light-cone
formalism takes into account effects from the relative motion of the $Q$
and $\overline Q$ in the quarkonium, which is neglected at leading order
in NRQCD. However, it is not clear that the model wave functions that are
used in these calculations are good approximations to the true
quarkonium light-cone wave functions.

There are also Belle results on inclusive double-charmonium production.
For the ratio $R_{J/\psi}=\sigma(e^+e^-\to J/\psi+c\overline c)
/\sigma(e^+e^-\to J/\psi+X)$, the most recent Belle analysis yields
$R_{J/\psi}=0.82\pm 0.15\pm 0.14$, with $R_{J/\psi}>0.48~\hbox{(90\%
confidence level)}$.\cite{belle-eps2003} Predictions based on NRQCD
factorization give $R_{J/\psi}\approx 0.1$ (Ref.~\refcite{YR}). In the
case of the absolute cross section for $J/\psi+c\overline c$ production,
the Belle result of 0.6--1.1~pb (Ref.~\refcite{Abe:2002rb}) disagrees
with the prediction of 0.10--0.15 pb (Ref.~\refcite{YR}) by almost an
order of magnitude. This prediction is based only on the color-singlet
contribution. However, corrections of higher order in $v$, including
color-octet contributions, are not expected to be large. Neither are
corrections of higher order in $\alpha_s$. It is difficult to see how
any perturbative calculation could give a value for $R_{J/\psi}$ as
large as 80\%.

\section{Summary}

The effective field theory NRQCD is a convenient formalism for
separating physics at the scale of the heavy-quark mass from physics at
the scale of quarkonium bound-state dynamics. The NRQCD factorization
approach provides a systematic method for calculating quarkonium decay
and production rates as a double expansion in powers of $\alpha_s$ and
$v$. NRQCD factorization for production rates relies upon
hard-scattering factorization and has not yet been established. The
NRQCD factorization approach has enjoyed a number of successes in
inclusive $P$-wave quarkonium decays, quarkonium production at the
Tevatron, $\gamma\gamma\rightarrow J/\psi +X$ at LEP, and quarkonium
production in DIS at HERA. Other processes, including production of
polarized quarkonium at the Tevatron, inelastic quarkonium
photoproduction at HERA, and double $c\bar c$ production at Belle and
BaBar are more problematic and point to the fact that our theoretical
understanding of quarkonium production is still incomplete.
%
%
%
%
\section*{Acknowledgements}

Work in the High Energy Physics Division at Argonne National
Laboratory is supported by the U.~S.~Department of Energy, Division of
High Energy Physics, under Contract No.~W-31-109-ENG-38.
%
%
%
%


\end{document}